\documentclass[aps,pre,twocolumn,twoside,preprintnumbers,amsmath,amssymb,longbibliography]{revtex4-1}

\usepackage{graphicx}
\usepackage{dcolumn}
\usepackage{bm}
\usepackage{amsmath,amsfonts,amssymb}
\usepackage{amssymb}
\usepackage{bbm}
\usepackage{mathrsfs}
\usepackage[sans]{dsfont}

\usepackage{amsmath,amsfonts,amssymb}
\usepackage[usenames]{color}
\usepackage[bookmarks, colorlinks=true, plainpages = false, citecolor = blue, linkcolor = blue, urlcolor = blue, filecolor = blue]{hyperref}

\usepackage{enumerate}

\newcommand{\gd}{\ensuremath{\dot\gamma}}

\newcommand{\etal}{{ \it et al.~}}
\newcommand{\cf}{{ \it cf.~}}
\newcommand{\sxy}{\ensuremath{\sigma}}
\newcommand{\yield}{\ensuremath{\sigma_{y}}}

\newcommand{\jamm}{\ensuremath{\phi_{J}}}

\newcommand{\beq}{\begin{equation}}
\newcommand{\eeq}{\end{equation}}

\newcommand{\mycolor}{\color{Black}}

\newcommand{\mc}{\color{Black}}

\begin{document}



\title{Characterizing the Nature of the Yielding Transition}

\author{S. H. E. Rahbari$^1$} \email{habib.rahbari@gmail.com}
\author{J. Vollmer$^2$}
\author{Hyunggyu Park$^1$}

\affiliation{$^1$ School of Physics, Korea Institute for Advanced
  Study, Seoul 130-722, South Korea}

\affiliation{$^2$ Faculty of Mathematics, Georg-August
  Univ. Göttingen, 37077 Göttingen, Germany}

\begin{abstract}


  Particulate matter, such as foams, emulsions, and granular
  materials, attain rigidity in a dense regime: the rigid phase can
  yield when a threshold force is applied. The rigidity transition in
  particulate matter exhibits {\it bona fide } scaling behavior near
  the transition point. {However, a precise determination of exponents
    describing the rigidity transition has raised much controversy.}
  Here, we pinpoint the causes of the controversies. We then establish
  a conceptual framework to quantify the critical nature of the
  yielding transition. Our results demonstrate that there is a
  spectrum of possible values for the critical exponents for which,
  without a robust framework, one cannot distinguish the genuine
  values of the exponents. Our approach is two-fold: ({\it i}) a
  precise determination of the transition density using rheological
  measurements and ({\it ii}) a matching rule that selects the
  critical exponents and rules out all other possibilities from the
  spectrum. This enables us to determine exponents with unprecedented
  accuracy and resolve the long-standing controversy over exponents of
  jamming. The generality of the approach paves the way to quantify
  the critical nature of many other types of rheological phase
  transitions such as those in oscillatory shearing.\\

\end{abstract}

\maketitle

\section{Introduction}

Yield stress materials such as toothpaste, hair gel, mayonnaise, and
cement, are ubiquitous. These materials are used in pharmaceutical and
cosmetics manufacturing, as well as the oil, concrete, and food
industries~\cite{bonn_2017}. Because of their wide applicability in
everyday life, a quantitative description of their rheological
behavior is pivotal. The physical origin of the yield stress depends on
the microscopic details of the system and can be classified into three
main categories: dynamic arrest in Brownian suspensions known as the
glass transition~\cite{petekidis_2004}, mechanical (meta)stability in
athermal systems or jamming~\cite{liu_1998, paredes_2013}, and
attractive interactions~\cite{trappe_2001, rahbari_2013}. Thixotropic
yield stress fluids~\cite{bonn_2017}, which exhibit memory effects and
a bifurcation in the viscosity, are outside the focus of the
current study. 

The relation between shear stress $\sigma$ and shear
rate $\gd$, known as a flow curve in a yield stress material, can be
described as a Herschel-Bulkley (HB) relation:
\begin{equation}
  \sigma = \sigma_y + K \gd^\Delta,
  \label{eq:HB}
\end{equation}
where $\sigma_y$ is the yield-stress and $\Delta<1$ is the shear
thinning exponent. In contrast, a simple Newtonian fluid is described
by a single parameter, namely the shear viscosity $\eta = \sigma /
\gd$. As a result of the threshold $\sigma_y$, the viscosity of a yield
stress material diverges for $\gd \to 0$. However, Barnes and
Walters~\cite{barnes_1985} demonstrated that carbopol microgels have
finite viscosity at small shear rates and raised a historical debate
over the existence of the yield stress. Two decades later, M\"oller \etal
~\cite{moeller_2009} repeated the same experiment and showed that
those measurements at low shear stresses never reached a stationary
state and that the apparent finite viscosity was an artifact of the
measurement.\\

A consensus regarding the existence of the yield stress has
emerged. However, the technical difficulties of its measurements remain a
challenge. Despite these advances in the understanding of the yield
stress, a description of the non-linear flow curves in the fluid state
remains an open problem. In the traditional approach, the shear-thinning
exponent $\Delta$ is obtained by a power law fit to $\sigma -
\sigma_y$ versus $\gd$. However, recent numerical simulations showed
that $\sigma - \sigma_y$ vs $\gd$ exhibits two distinct scaling
regimes described by two different exponents, $\Delta$ and
$\Delta^\prime$, for small and large shear rates~\cite{olsson_2011,
  lerner_2012, kawasaki_2015}, respectively. As a result, fitting a
HB-type relation to such data will be prone to pitfalls
due to a bias towards larger shear rates, which in turn will give rise
to a misleading quantification of the flow curves.\\

The problem becomes even more dramatic for the case of matter with
granularity~\cite{schall_2010}. Soft particulate matter, such as gels
and emulsions, flow freely in the dilute regime and attain yield stress
above a threshold density $\jamm$ in the dense regime. This yielding
transition exhibits a rich class of scaling behavior of the flow
curves described by critical exponents (we will give a brief overview
about different scaling regimes of the rigidity transition in the next
section). Despite many efforts by different
groups~\cite{hatano_2008, otsuki_2009_2, hatano_2010, otsuki_2012,
  degiuli_2015, vagberg_2016}, a precise determination of the critical
exponents remains disputed.


Here, we establish a conceptual framework for the scaling quantification
of the flow curves of a wide range of yield stress materials. We
resolve the long-standing dispute over exponents of the rigidity
transition.

\section{Rigidity transition: a bird's eye view}
\label{Sec:bird_eye}

Depending on the shear rate and packing fraction, soft frictionless
spheres display a rich phenomenology of distinct rheological
regimes. This makes soft frictionless spheres {\it Drosophila} of
particulate matter.\\

In the dilute regime of particulate materials, flow curves at small
shear rates are given by $\sxy \propto \gd ^n$ where $n = 1$ and $2$
for Newtonian and Bagnoldian scalings, respectively. Because soft
particles barely deform at small shear rates, $\sxy \propto \gd ^n$
corresponds to the so-called hard-core limit. The exponent $n$ has
been shown to depend on the Reynolds number of the system such that
for overdamped systems the Newtonian regime (non-inertial) is
recovered and for $n=2$ the system must be under-damped
(inertial)~\cite{vaagberg_2014_2}. The transport coefficient, which is
given by shear viscosity $\eta = \sxy / \gd ^ n$, at $\gd \to 0$,
depends only on the packing fraction $\phi$ and diverges upon
approaching the jamming density $\eta \propto |\delta \phi|
^{-\beta}$, where $\delta\phi = \phi - \jamm$ is the distance from
jamming. The exponent $\beta$ also characterizes the hard-core limit
of the system. Accordingly, this exponent must be independent from the
microscopic details of the system~\cite{olsson_2012, vagberg_2014}. At
$\phi = \jamm$, the system exhibits pure power-law rheology $\sxy
\propto \gd ^ q$ with $q< 1$ as the critical shear-thinning
exponent. In the soft core regime $\phi > \jamm$, the system displays
threshold rheology and flow curves that may be described by the HB
model given by Eq.~\ref{eq:HB}. In this model, the shear-thinning
exponent $\Delta$ is shown to be related to the behavior of the system
in the hard-core limit at $\phi < \jamm$ and thus to the exponent
$\beta$~\cite{olsson_2012}. The yield stress also scales with the
distance from jamming $\yield \propto \delta \phi ^ y$.

As one can see, upon
approaching the jamming point, the rheology changes dramatically due
to the collective behavior of particles~\cite{radjai_2002}. Consequently,
the rheology can no longer be described by trivial exponents such as
$n = 1$ or $2$ and thus the system becomes shear-thinning with a
non-trivial scaling dimension $q < 1$. This is a signature of a
growing length scale in the system~~\cite{olsson_2011, olsson_2015,
  nordstrom_2010}, which is the hallmark of critical phenomena. Even
though this system is non-equilibrium and athermal, Olsson and
Teitel~\cite{olsson_2007} used renormalization group
formalism~\cite{kardar_2007} of equilibrium phase transitions to
capture the critical nature of this dynamic transition. The jamming
point at $\delta\phi = 0, \gd \to 0, T = 0,$ and $L \to \infty$ is a
genuine dynamic critical point.

Altogether, any of the above scaling limits can be retrieved
by choosing appropriate limits of a scaling function $\mathcal{F}_0$
and an arbitrary length scale $b$ in the following scaling ansatz:
(derivation given in Appendix \hyperref[append:scaling_ansatz]{B}):
\begin{align}\label{eq:scaling-ansatz}
  \sigma( \delta\phi , \gd, L, w ) = b^{-y/\nu} \; \mathcal{F}_0(
  \delta\phi \: b^{1/\nu} , \gd \: b^z, L^{-1} \: b, w \: b^{-\omega}
  ),
\end{align}
where $\mathcal{F}_0$ is a homogeneous scaling function, $L$ is the
system size, and $w$ is an auxiliary variable. This scaling ansatz is
traditionally used to find relations between different exponents.
Inserting $b = \gd^{-1/z}$ for $L\to\infty$ in
Eq.~\ref{eq:scaling-ansatz}, we arrive at:
\begin{equation}
  \sxy = \gd ^ q \mathcal{F}_1\left( \frac{\delta \phi}{ \gd ^
    {q/y}}\right),
  \label{eq:lead_scal}
\end{equation}
where $q = y / (z\nu)$. Here, we assume proximity of the critical
point where the auxiliary variable $w$ can be neglected. 

The immediate outcome of Eq.~\ref{eq:lead_scal} is that all the data
must collapse into a master curve when plotted $ \sxy / \gd ^ q$ vs
${\delta \phi} / { \gd ^ {q/y}}$, providing that three free parameters
$q$, $y$, and $\jamm$, are fine tuned. Notably, in the early stage of
this topic, this method, \i.e., collapse of the data, has been
extensively used by many authors to estimate $q$, $y$, and
$\jamm$~\cite{olsson_2007, hatano_2008, hatano_2010, otsuki_2012,
  hayakawa_2013}. A summary of the existing predictions for these
exponents is given in Tab.~\ref{tab:exponents-from-literature}. These
reports were not conclusive because of the large range of reported
exponents and critical densities. The reason for this was because the
quality of the collapses were judged based on the visual appeal of the
plots. Later, Olsson and Teitel used a quantitative method to compute
the quality of the collapses. The method was based on ({\it i})
exponential parametrization of the scaling function
$\mathcal{F}_1(x)=\exp\left(\sum_{n=0}^5a_nx^n\right)$ and ({\it ii})
going into unprecedented small shear rates down to $10^{-8}$ in the
dimensionless scale~\cite{olsson_2011, vagberg_2016}. However, the
expansion of $\mathcal{F}_1(x)$ may be prevented because, as $x\to 0$,
$\mathcal{F}_1(x)$ may not be analytic. Also, for reasons that we
describe in the next paragraph, going into shear rates as small as
$\gd\approx 10^{-8}$ contaminates the scaling behavior.\\

\begin{table}
  \[\begin{array}{llll}
  \hline \hline \text{authors} & y & q & \omega/z\\
  \hline\\[-2mm]
  \text{Otsuki \& Hayakawa (theory)~\cite{otsuki_2009_2}}  & 1   & 2/5  & - \\[1mm]
  \text{Hatano~\cite{hatano_2008}}                         & 1.2 & 0.63 & - \\[1mm]
  \text{Hatano~\cite{hatano_2010}}                         & 1.5 & 0.6  & - \\[1mm]
  \text{Otsuki \& Hayakawa (simul.)~\cite{otsuki_2012}}    & 1.09 & 0.46 & -\\[1mm]
  \text{DeGiuli~\etal~\cite{degiuli_2015}}                 & 1 & 0.3 & 0.3  \\[1mm]
  \text{Vagberg~\etal~\cite{vagberg_2016}}                 & 1.15(5) & 0.38(5) & 0.35(7) \\[1mm]
  \text{Goodrich\etal~\cite{goodrich_2016}}                & 1 & - &
  - \\ \hline \hline
  \end{array}\]
  \caption{ Critical exponents reported by different authors. As the
    data get closer to the critical point, exponents systematically
    change (for a comprehensive discussion, \cf~\cite{vagberg_2016}).
    \label{tab:exponents-from-literature}}
\end{table}
%

It is well known that in the jammed state a sheared particulate system
exhibits shear localizations, also known as
shear-transformation-zones~\cite{maloney_2006,
  bouchbinder_2007}. These stress anomalies relax through long-range
system-wide avalanches. Each avalanche can trigger other active zones
that will in turn result in a domino of plastic events and
relaxations. At very small shear rates, these avalanches are globally
correlated and poise the system into an {\em effective } critical
state~\cite{lemaitre_2009, hentschel_2010}. This results in scale-free
distributions of avalanches with exponents that are generally smaller
than 2~\cite{hatano_2015, nicolas_2017}. To obtain a flow curve
$\sxy\left(\gd\right)$, one should perform time averaging for shear
stress over the time series. However, due to the scale-free
distribution of avalanches with the aforementioned range of exponents,
the first and second moments of the shear stress cannot be
well-defined. Consequently, the time averaged shear stress at very
small shear rates possesses error bars that are as large as the
average values.\\

 To avoid the above problems, we describe a general
 framework that requires neither data collapse nor expansion of the
 scaling function. Additionally, measurements are performed outside the avalanche
 region, \i.e., not at very small shear rates. These simplify the problem
 dramatically and enable us to resolve the controversy over
 exponents. Our approach is two-fold: first, in
 Sec.~\ref{sec:hunt_phiJ}, we describe how we nail down the critical
 density. Second, in Sec.~\ref{sec:hunt_exponents}, we present our
 matching rule that selects critical exponents from a wide spectrum of
 possible values. \\

\section{Hunt for $\jamm$}
\label{sec:hunt_phiJ}

Precise determination of the critical exponents strongly depends on
whether the critical density $\jamm$ is accurately determined. In this
section, we explain how we nail down the transition density $\jamm$
using rheological data. To achieve this goal, we define successive slopes of
the flow curves $m$ as:
\begin{equation}
  m = \frac {d\ln\sxy}{d\ln\gd},
  \label{Eq:eff_slope_def}
\end{equation}
where $d$ stands for the derivative. This can be easily calculated from
Eq.~\ref{eq:lead_scal}:
\begin{equation}
  m = q - \frac{q}{y} \frac{\delta \phi}{ \gd ^ {q/y}}
  \frac{\mathcal{F}_1^\prime(x)}{\mathcal{F}_1(x)},
  \label{Eq:eff_slope}
\end{equation}
where $x = {\delta \phi} / { \gd ^ {q/y}}$, $\mathcal{F}_1^\prime(x) =
d\mathcal{F}_1(x)/dx$.\\

Eq.~\ref{Eq:eff_slope} provides an immediate prediction: if one plots
$m$ vs $\gd$ for different packing fractions, exactly at jamming
density $\delta \phi = 0$, the successive slope for all shear rates
will be equal to the critical shear-thinning exponent $m = q$. For
$\delta \phi > 0$, the successive slope converges to $m = q$ at large
shear rates and deviates from that value for $\gd \to 0$ according to
$\gd^{-q/y}$. Similar behavior is expected for $\delta \phi < 0$ with
an opposite curvature.\\

This provides a simple recipe to compute $\jamm$: the critical density
is given by a horizontal line of the $m-\gd$ dependence that distinguishes
off-critical densities with opposite curvatures. However, it is
practically impossible to recover a straight horizontal line for $m$
at $\jamm$ in the critical region of $\gd\to 0$. This is due to
elasto-plastic critical fluctuations near the critical point, which we
mentioned in Sec.~\ref{Sec:bird_eye}.\\

The remedy for this problem is to stay away from the region where the
successive slope displays huge fluctuations. In such a regime,
correction-to-scaling must be taken into account. From
Eq.~\ref{eq:scaling-ansatz}, the leading correction-to-scaling term at
$\phi=\jamm$ reads:
\begin{equation}
  \sigma = \gd^q\left( c_1 + c_2 \gd ^{\omega/z}\right),
  \label{eq:stress_gd_corr_to_scal}
\end{equation}
where $c_1$ and $c_2$ are constants and $\omega/z$ is the leading
correction-to-scaling exponent (see Appendix
\hyperref[append:scaling_ansatz]{B} for derivation). For off-critical
densities $\phi\ne\jamm$, an extra term proportional to $\delta\phi$
must be added to Eq.~\ref{eq:stress_gd_corr_to_scal}. This term again
has an inverse algebraic dependence on $\gd$ similar to that in
Eq.~\ref{Eq:eff_slope}. One can easily calculate the corresponding
successive slope of Eq.~\ref{eq:stress_gd_corr_to_scal} as:
\begin{equation}
  m = q + k \gd ^{\omega/z},
  \label{Eq:succ_slope_corr}
\end{equation}
where $q$ is the asymptotic exponent and $k$ is a constant. This shows
the behavior of the successive slopes at $\jamm$, which distinguishes that
of off-critical densities with opposite curvatures.\\

Now let us calculate the asymptotic values of the successive slopes for
different densities at $\gd\to 0$. For $\delta\phi<0$, $\sigma\propto
\gd^n$, which results in $m=n$. At $\delta\phi=0$, $\sigma\propto
\gd^q$, then $m=q$. For $\delta\phi>0$, the yield stress emerges, which
amounts to a dependence $\propto\gd^0$ and thus $m=0$. In summary:
\begin{equation}
  \lim_{\gd\to 0} m = \left\{
  \begin{array}{lr}
    n &   \phi < \jamm \\
    q &   \phi = \jamm \\
    0 &   \phi > \jamm. \\
  \end{array}
  \right.
  \label{braket:m_asympt}
\end{equation}

\begin{figure}[b]
  \centering
  \hfill\includegraphics[width=0.5\textwidth]{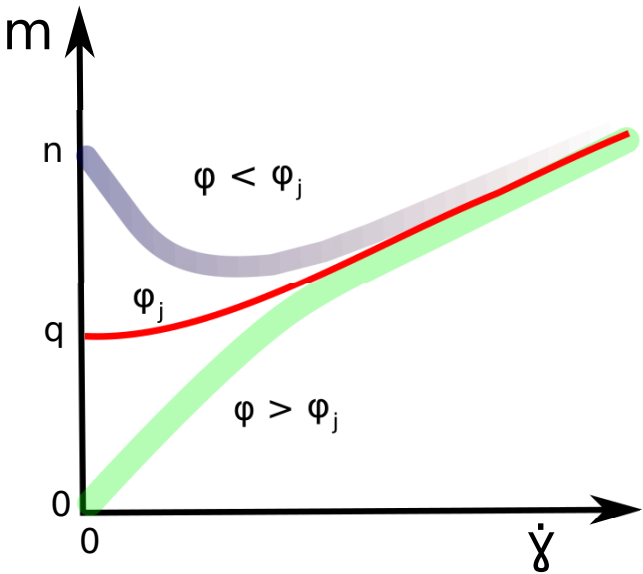}
  \hfill\\
  \caption[]{{\bf Schematics of successive slope.} A schematic figure
    shows the successive slope vs shear rate in a semi-log
    scale. According to Eq.~\ref{Eq:eff_slope}, curves at sub- and
    super-critical densities have opposite curvatures. In the
    asymptotic limit $\gd \to 0$, all curves converge to the
    asymptotic exponents given by Eq.~\ref{braket:m_asympt}. The
    curve corresponding to $\jamm$ {\em is the only curve} that does
    not bend upward or downward and whose offset is equal to the
    non-trivial critical exponent $q$.
    \label{fig:schemataic_succ_slope}}
\end{figure}

We summarize the behavior of the successive slope of flow curves in a
schematic diagram in Fig.~\ref{fig:schemataic_succ_slope}. This
diagram demonstrates the simplicity behind our framework to find
$\jamm$. In a semi-log plot of $m$ vs \gd, all of the sub- and
super-critical densities curve in opposite directions, except
at $\jamm$.\\


In our strategy to find $\jamm$, we first obtain flow curves for an
intermediate system size. We mark the range of densities where the curvature
of the successive slopes changes. We then zoom into the region by
simulating a larger system size and nail down $\jamm$. Finally, we
check whether our estimated $\jamm$ is robust against finite-size
effects.\\

We perform extensive large-scale two dimensional molecular dynamics
simulations of frictionless disks in a simple shear flow. {\mc In our
  simulations, we dissipate the normal component of the relative
  velocity of colliding particles. This dissipation law leads to
  Bagnoldian scaling in the dilute regime. The Newtonian regime is
  recovered when the transverse component of the relative velocity is
  dissipated~\cite{vagberg_2014}. This regime is not explored in this
  work}. Further details of the simulations are given in Appendix
\hyperref[append:simulations]{A}. In Fig.~\ref{fig:succ_slope_lead},
we display the successive slope $m$ vs shear rate $\gd$ for different
packing fractions $\phi$ for a system of intermediate size
$L=100$. The curvature of the curves changes in the range between
$\phi = 0.843$ and $0.844$. This determines the window for $\jamm$. We
will zoom into this region to determine $\jamm$ with a higher
resolution and larger system sizes. All of the curves corresponding to
different packing fractions show a tendency to converge at large shear
rates. This is in accord with the predictions by
Eq.~\ref{Eq:eff_slope}. One can see that upon decreasing the shear
rate, the far-top curves show a tendency to converge towards the value
of the asymptotic exponent $n = 2$ and the far-bottom curves to
$0$. This is again in accord with the prediction by
Eq.~\ref{braket:m_asympt}. A dashed line shows an estimation for the
value of $q=0.6$. We note that this line tends towards smaller values
upon increasing $L$. For $L\ge 200$, the estimated value of $q$ does
not change. We note that for $\gd<10^{-6}$, the successive slope in
the critical range of densities displays giant fluctuations
reminiscent of critical fluctuations. We observe these fluctuations
for systems of larger spatial extents for $\gd<10^{-6}$. Therefore, in
the rest of the paper, we do not consider data with $\gd<10^{-6}$ in
our analysis. { To summarize the results for $L = 100$, the crude
  estimation for the transition density is $\jamm\approx
  0.84335\pm0.00035$. The na\"ive estimation for the critical exponent
  is $ q\approx 0.6$. Next, we will zoom into the critical region with
  substantially larger system sizes to find $\jamm$.}\\

\begin{figure}[b]
  \centering
  \hfill\includegraphics[width=0.5\textwidth]{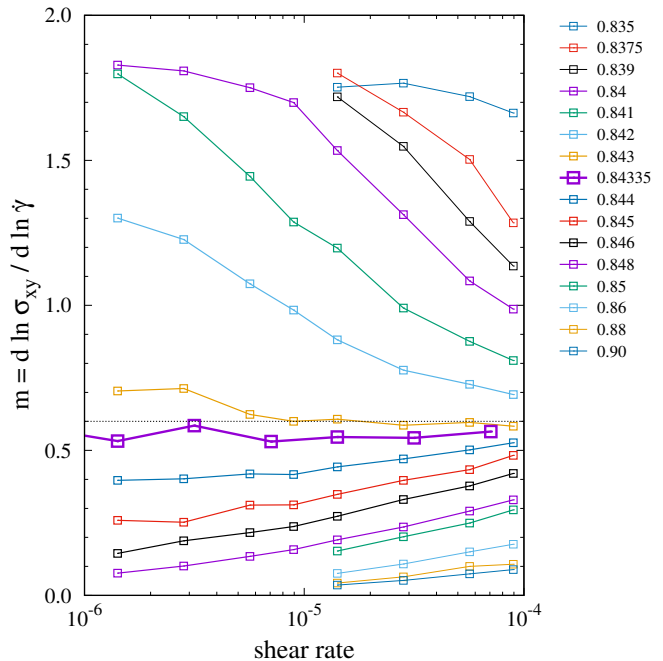}
  \hfill\\
  \caption[]{{\bf Successive slope. } Successive slope $m$ vs shear
    rate $\gd$ for different $\phi$. The value of each $\phi$ is given in
    the legend. The number of particles varies from $N = 7183$ to $7742$ for
    $\phi = 0.835$ and $0.90$, respectively. The spatial extent of the
    system is $L=100$. The curvature of the curves changes in the range of
    $\phi = 0.843$ and $0.844$. This marks the critical window for
    $\jamm$ and the range for refined measurements with a much larger
    resolution on $\phi$ for substantially larger system sizes.
    \label{fig:succ_slope_lead}}
\end{figure}




{\mycolor According to elasticity theory, shear stress and pressure
  are both components of a single entity known as the stress tensor.
  Different components of the stress tensor provide information about
  momentum transfer in different directions into/along imaginary
  surfaces in the system~\cite{landau_1986}. However, whether the
  shear stress and pressure scale equivalently with shear rate is not
  at all an obvious fact. According to Peynneau and
  Roux~\cite{peyneau_2008_2} and more recently by Baity
  \etal~\cite{baity_2017} a finite stress anisotropy, $\delta p
  \propto p_{xx} - p_{yy}$, gives rise to a small rotation of
  principal axes of of the stress tensor from those given by the
  strain tensor. This gives rise to distinct scaling of the shear
  stress and pressure when there is a stress anisotropy in the system;
  this usually happens at high shear rates. However, the stress
  anisotropy is negligible at small shear rates near
  jamming~\cite{vagberg_2017}. This is also confirmed by our
  results. Thus, it is a widely accepted fact that the asymptotic
  scaling of the shear stress and pressure are equivalent. This
  assumption has been adopted by many recent studies,
  \cf~\cite{vagberg_2016}. More recently, Suzuki and Hayakawa provided
  a rigorous derivation of this based on a $\mu $-$J$
  rheology~\cite{suzuki_2017}. We will use this assumption in the next
  section to nail down the critical exponent $q$.\\ }

\begin{figure*}
  \[
  \includegraphics{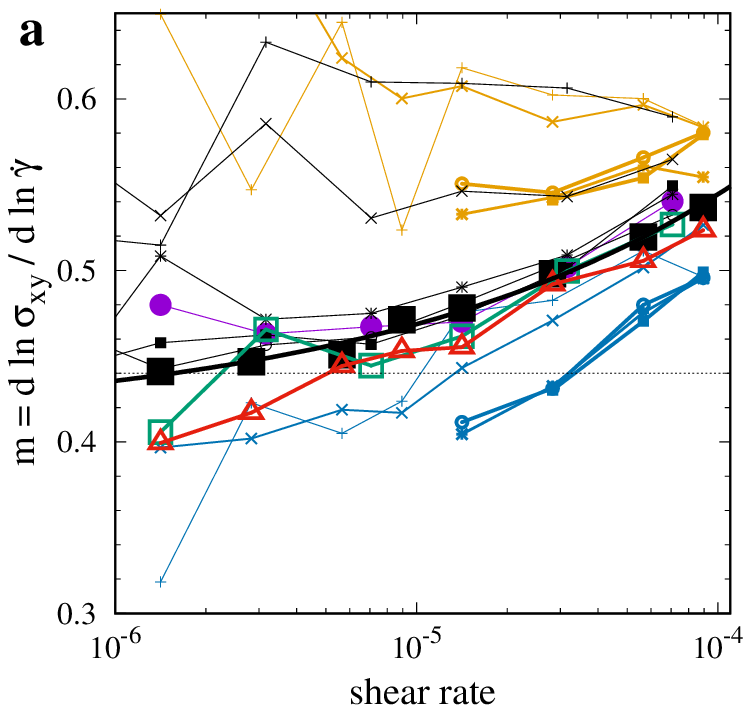} \qquad
  \includegraphics{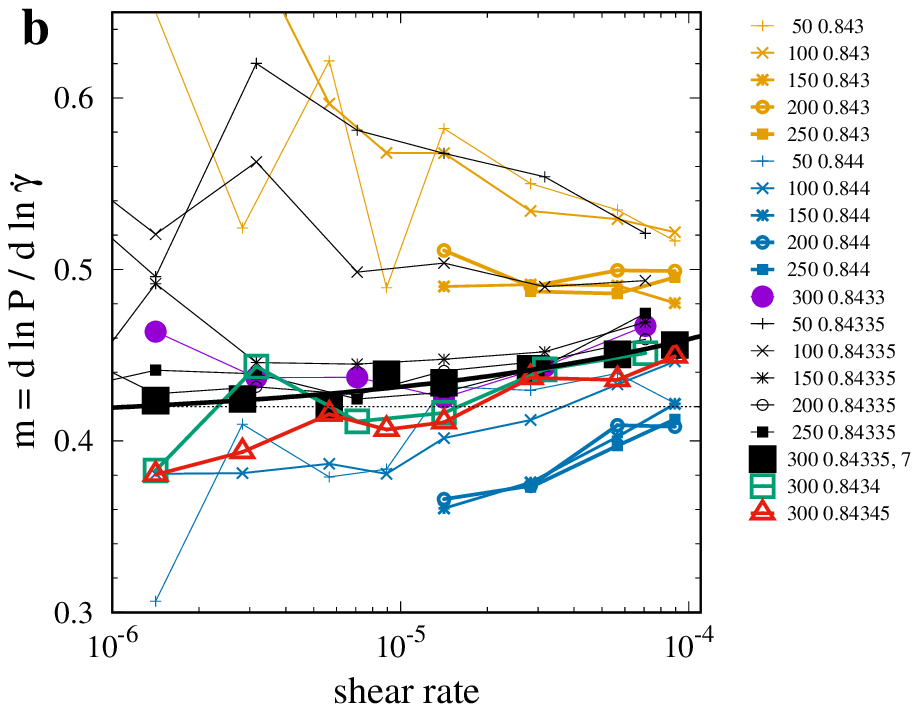}
  \]
  \hfill\\
  \caption[]{{\bf Successive slope of refined measurements.}
    Successive slope versus shear rate for ({\bf a}) shear stress and
    ({\bf b}) pressure. The critical density is found $\jamm =
    0.84335$. We observe a strong system size dependence for $L\leq
    200$. For $L\ge 250$, different system sizes are on top of each
    other. In this density, for $L = 300$, the system consists of $N =
    65397$ particles. The successive slopes at slightly above and
    below $\jamm$ bend in opposite directions. A stronger
    correction-to-scaling is found for the shear stress $\sxy$.
    \label{fig:eff_slope_corr}}
\end{figure*}

We display refined measurements in Fig.~\ref{fig:eff_slope_corr} for
different system sizes up to $L = 300$. Panel {\bf a} and {\bf b}
refer to the successive slopes of the shear stress and pressure,
respectively. One can see that for all densities, there is a strong
system size dependence for $L<200$. For $L\ge 200$, the successive
slopes are on top of each other for all densities. The curves at
$\phi=0.843$ and $0.844$ clearly have opposite curvatures for all
system sizes. We zoom into this region to find the critical
density. Filled squares correspond to $\phi=0.84335$ and
$L=300$. These data are averaged over $7$ different ensembles. The
rest of the data are obtained from a single realization. For $L=300$,
a closer inspection of data at $\phi=0.8433$ and $0.8434$ reveals
their opposite curvatures. The $\phi=0.84345$ line is curved down
similar to that at $\phi=0.8434$.  Therefore, these are off-critical
densities. However, one can clearly see that $\phi = 0.84335$ (filled
squares) is the cross-over density where the curvature
changes. Therefore, we conclude $\jamm=0.84335\pm
0.00005$. Interestingly, our estimated density within error bars
agrees with that of Heussinger \etal~\cite{heussinger_2009} and
Vagberg \etal~\cite{vagberg_2016}. {\mycolor A closer inspection of
  the successive slope of shear stress $\sxy$ (panel {\bf a}) and
  pressure $p$ (panel {\bf b}) reveals a stronger
  corrections-to-scaling of the shear stress. Here, stronger
  corrections-to-scaling means a larger amplitude of the scaling
  function of Eq.~\ref{Eq:succ_slope_corr}. However, as we have
  mentioned in the previous paragraph the asymptotic exponents must be
  equivalent for both pressure and shear stress. Interestingly, a
  stronger corrections-to-scaling of shear stress has been reported by
  other authors ~\cite{olsson_2011, vagberg_2016}.}\\

{ One can see that $\jamm$ does not have a strong
  dependence on the system size. However, the asymptotic exponent
  changes continuously from $0.6$ to approximately $0.4$ by
  increasing the system size from $L=50$ to $300$,
  respectively. Estimation of $q$ for $L=300$ is not straightforward
  because of the complexity of the scaling function for large system
  sizes, \i.e., the dependence of $m$ to $\gd$. In the next section, we
  describe a systematic method to nail down the critical exponents.}

\section{Hunt for exponents}
\label{sec:hunt_exponents}

In this section, we describe how we nail down the critical
exponents. The easiest way to find critical exponents is to obtain
them via fitting Eq.~\ref{Eq:succ_slope_corr} to the successive slope
curve at $\jamm$ in Fig.~\ref{fig:eff_slope_corr} using $q$, $k$, and
$\omega/z$ as free fitting parameters. We call this a {\em blind}
fitting. Notably, a $3$-parameter fitting corresponds to the optimization
of a residual function in a $3+1$ dimensional space. This function is
rugged and has many local basins. Each fitting algorithm/software
will find one such local minimum. This will cause a zoo of different
values for the exponents due to the rugged nature of the residual
function.\\

\begin{figure*}
  \[
  \hfill\includegraphics[width=0.5\textwidth]{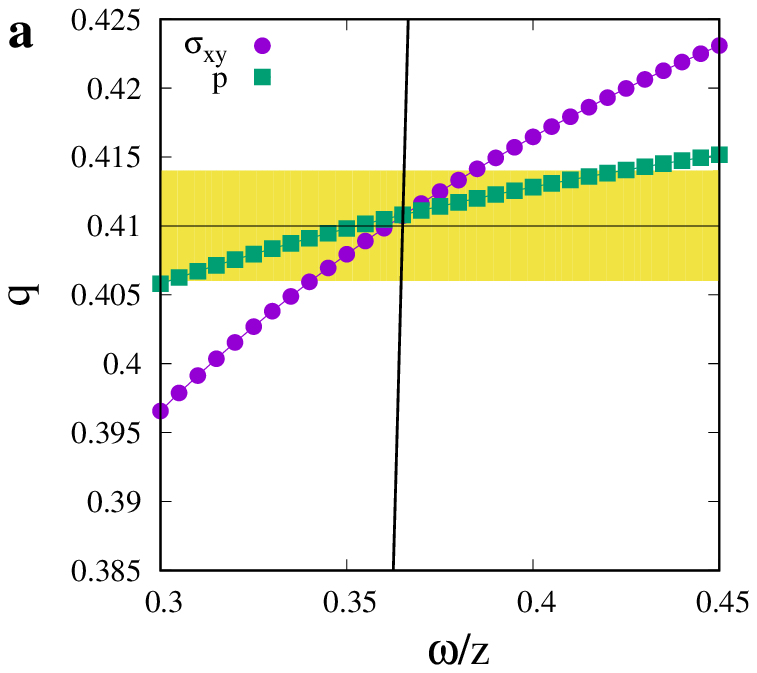}
  \qquad
  \hfill\includegraphics[width=0.5\textwidth]{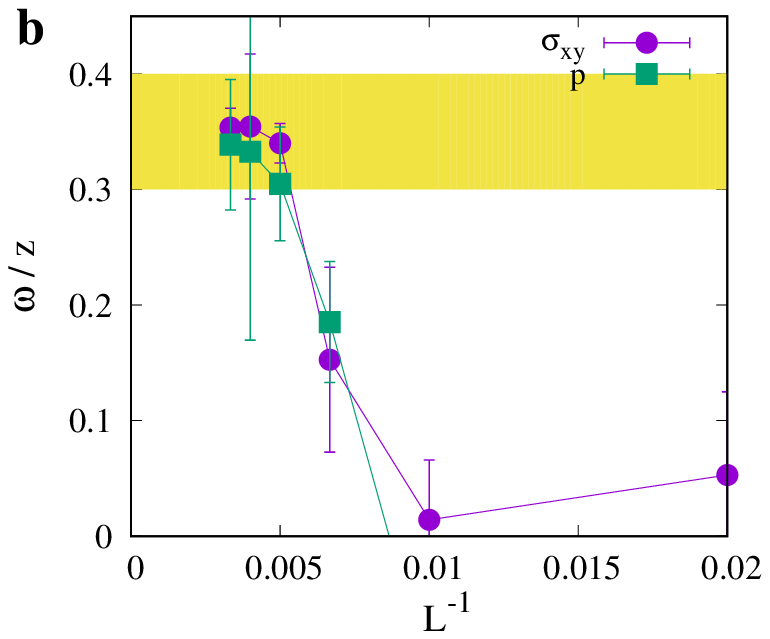}
  \]
  \hfill\\
  \caption[]{{\bf Matching rule.} ({\bf a}) Contour lines of $q$ and
    $\omega/z$ for $p$ and $\sxy$. The matching rule selects critical
    exponents where the contour lines of $p$ and $\sxy$
    intersect. ({\bf b}) We hold fixed $q\left(L=\infty\right)=0.41$
    and obtain correction-to-scaling exponent $\omega/z$ via fitting
    through Eq.~\ref{Eq:succ_slope_corr}. The obtained value of
    $\omega/z\left(L\right)$ is plotted against $L^{-1}$. One can see
    that $\omega/z$ saturates for large system sizes to
    $\omega/z\left(L=\infty\right)=0.35$.
    \label{fig:match_rule}}
\end{figure*}

To avoid fitting artifacts, we hold the correction-to-scaling exponent
$\omega/z$ fixed and obtain the asymptotic exponent $q$ via fitting a
linear function to $m$ versus $\gd^{\omega/z}$. We vary $\omega/z$ in
a range between $0.3$ and $0.5$, and we record the corresponding
$q$. The contour lines of the fits are given in
Fig~\ref{fig:match_rule}-a for both shear stress and pressure. Each
contour line represents all possible outcomes of $q$ and $\omega/z$
via a $3$-parameter blind fitting. Each point on the contour lines
corresponds to one basin. Now, the crucial question becomes about which point
on the contour line can be considered the corresponding point for
correct exponents.\\

As we previously noted, pressure and shear stress can have different
scaling functions; however, asymptotic critical exponents must
strictly be equivalent. This provides a {\em matching rule}, which
allows us to pick up the correct exponents based on the crossing point
of the contour lines of $p$ and $\sxy$. Fig.~\ref{fig:match_rule}-a
demonstrates that such a matching point really does exist, and we read
exponents $q=0.41$ and $\omega/z=0.365$ for both $p$ and $\sxy$. We
note that within error bars, the crossing point gives the same $q$ for
$L\ge 200$. However, $\omega/z$ is not stable. Therefore, we perform
the finite-size scaling analysis for $\omega/z$ with fixed
$q\left(L=\infty\right)=0.41$ via
\begin{equation}
  m\left(L\right) - q\left(L=\infty\right) \propto
  \gd^{\omega/z\left(L\right)}.
  \label{eq:omega_over_z_fss}
\end{equation}

We fit Eq.~\ref{eq:omega_over_z_fss} and obtain $\omega/z$ as a
function of $L$. We plot $\omega/z\left(L\right)$ vs $L^{-1}$ in
Fig.~\ref{fig:match_rule}-b. One can see that $\omega/z$ levels off at
$0.35$ for the largest system sizes. This gives us the asymptotic
value of the leading correction-to-scaling exponent
$\omega/z\left(L=\infty\right) = 0.35$. We summarized the values of
critical exponents in Tab.~\ref{Tab:fits}.\\

Having obtained both $q\left(L=\infty\right)$ and
$\omega/z\left(L=\infty\right)$, we arrive at our final vital
benchmark. We now hold $q$ and $\omega/z$ fixed to their asymptotic
$L=\infty$ values and fit Eq.~\ref{Eq:succ_slope_corr} to the data to obtain the amplitude $k$. The resulting curves are shown as
solid lines in both Fig.~\ref{fig:eff_slope_corr}-a and -b. We obtain
$k = 3.7$ and $1.36$ for shear stress and pressure,
respectively. Since $k$ is the amplitude of the leading
correction-to-scaling term, which is supposed to be a small term, $k$
must be $\mathcal{O}(1)$. This dramatically depends on the window of
$\gd$. If this window is far from the critical region, then the next terms
in the correction-to-scaling must be considered. Moreover, for such
cases where the window of $\gd$ is far from the critical region and
only the leading correction-to-scaling is considered, the obtained
value of $k$ becomes too small or too large. Here, we see that we
arrive at conclusive values of $k\sim\mathcal{O}(1)$ for both $\sxy$
and $p$. This consistency check is crucial for the analysis and must
be carried out to examine the pre-assumptions for the
correction-to-scaling terms.\\

{ As a final note, the exponent $y$ describes how the yield stress
  $\sigma_y$ scales with distance to jamming $\delta\phi$. This
  exponent can be measured by simulations of pure isotropic
  compression, and no-shearing is required. It is well known that
  $y\simeq 1$~\cite{ohern_2003}.}

\begin{table*}[t]
\centering
\begin{tabular}{ p{7.5cm} p{7.5cm} p{1.5cm} }
 \hline\hline
 Exponent    & $q$         &  $\omega/z$     \\
 \hline
 $\sxy, p$   & $0.410(5)$  &  $0.35(5) $ \\
 \hline\hline
 \end{tabular}
\caption[]{{\bf Exponents.} Numerical values of critical exponents. }
\label{Tab:fits}
\end{table*}

\section{Discussion and conclusion}

Soft spheres flow freely in a dilute regime and become amorphous solid
in a dense regime. This accounts for a large range of phenomena such
as jamming and glass transition. Determinations of both the transition
density and exponents describing scaling near the transition point are
subjects of intense research. However, because of a lack of a general
framework, no consensus has yet emerged. Here, we close this debate by
presenting a framework to precisely compute the exponents of the
rigidity transition in soft spheres based on an accurate determination
of the transition density. Furthermore, we demonstrate that even
though the transition density can be uniquely determined, there is a
spectrum of different numerical values for critical exponents. Thanks
to isotropic {\mycolor asymptotic} scaling of the components of the
stress tensor, we introduce a matching rule that selects critical
exponents and rules out other possibilities. The matching rule
considers the intersection of contour lines of exponents of pressure
and shear stress. This allowed us to unambiguously determine the
{\mycolor asymptotic} critical exponent of the shear stress and
pressure. {\mycolor Having determined the asymptotic exponent $q$, we
  use finite-size scaling to determine the asymptotic value of
  exponent of the leading correction-to-scaling term $\omega/z$. We
  demonstrate that $\omega/z$ for both shear stress and pressure
  converges to the same value within the numerical uncertainty at the
  limit of large system sizes.} Two mean-field type calculations for
the exponents of the rigidity transition are proposed by
Otsuki-Hayakawa~\cite{otsuki_2009} and DeGiuli \etal
\cite{degiuli_2015}. Our results for exponents are closer to the
predictions by the former.\\

{\mycolor Noticeably, we recover inertial-Bagnold scaling $p,
  \sigma \propto \gd^2$ at $\gd\to 0$ below jamming. This is a
  direct result of the fact that our dissipation rule damps out the
  normal component of the relative velocity with respect to the
  contact point of two colliding particles. However, since in a shear
  flow the main contribution to the kinetic energy of particles comes
  from the tangential relative velocities of particles, after a
  collision particles maintain their motion due to the apparent
  inertia. This fact was first noted by Refs.~\cite{vagberg_2014,
    degiuli_2015}.\\

}

{ Even though the critical density is not strongly influenced by
  finite size effects in our analysis, we observe a strong dependence
  of the critical exponents on the system size. This is a crucial
  point that has been overlooked in many recent studies about glass
  transition and jamming. In these studies, extremely small system
  sizes, in the order of $10^3$ particles, are considered. Our results
  indicate that such small system sizes are strongly influenced by
  finite size effects.}\\

Our framework provides grounds for several immediate investigations
that will deepen our understanding of amorphous materials using
rheology as the main tool:


\begin{enumerate}[(i)]

\item { Critical exponents of a phase transition can be
  influenced by fluctuations and thus the dimensionality of the
  system. However, these exponents do not significantly change above a
  critical dimension, known as the upper critical dimension $d_{UC}$. Below
  this dimension, fluctuations are important. Above $d_{UC}$,
  fluctuations are washed out and critical exponents are equal to the
  mean-field exponents. The exact determination of $d_{UC}$ for the
  jamming transition has been a challenge: the absence of a mean field
  theory and the lack of a framework for the precise measurement of critical
  exponents can be considered as the main reasons. Many authors have suggested that $d_{UC} = 2$ and that logarithmic
  corrections-to-scaling are involved~\cite{ohern_2003, wyart_2005_2,
    goodrich_2012, goodrich_2014_2, goodrich_2016}. The main reason
  for this is that critical exponents appear to be the same for $d = 2$ and
  $3$. Collapse of the data has been used widely in these studies to
  measure critical exponents. Our general approach can be easily
  applied in accurately measuring critical exponents in three
  dimensions. Then, a comparison of critical exponents at $d=2$ and
  $3$ can resolve the controversy over the upper critical dimension
  for jamming. This will be a great step ahead in understanding the
  nature of the jamming transition.}

\item { Amorphous solids possess a complex free-energy
  landscape~\cite{charbonneau_2014}. As one increases density, an
  amorphous solid undergoes a sequence of transitions: glass
  transition, Gardner transition, and jamming. Annealing has been the
  essential method to investigate these transitions.
  Standard rheological techniques have been shown to be powerful tools
  to investigate complex properties of this energy landscape of
  amorphous materials~\cite{jin_2017}. We expect the generalization of
  our approach to shed light on and help formulate a general
  formalism to investigate other types of transitions in amorphous
  materials using rheology. }

\item An investigation of periodically driven colloidal suspensions
  provided remarkable insights into the nature of rheological phase
  transitions. In a dilute regime, these systems undergo a
  non-equilibrium phase transition into an absorbing state where
  particles self-organize themselves to prevent
  collisions~\cite{pine_2005, corte_2008}. In the dense regime,
  a yielding transition, which describes the onset of plastic deformation,
  has been shown to be a non-equilibrium phase transition from
  reversibility in an elastic regime into irreversibility in a plastic
  regime~\cite{nagamanasa_2014}. Nonetheless, the nature of the
  transition has been disputed, including whether it is a first-order or second-order
  transition~\cite{regev_2015, leishangthem_2017}, as well as whether the
  absorbing state transition belongs to the universality class of (conserved)
  directed percolation~\cite{nagamanasa_2014}. In
  both the dilute and dense regimes, divergences of time and length scales
  have been reported upon approaching a critical shearing
  strain. However, a precise determination of the universality classes of
  these non-equilibrium phase transitions have not been conclusive due
  to the lack of a comprehensive framework for measuring critical
  exponents from rheological experiments~\cite{tjhung_2015, pine_2005,
    corte_2008, nagamanasa_2014, jeanneret_2014}. Our formalism may
  shed light on resolving the dispute over the nature of rheological phase
  transitions in oscillatory shearing.

\end{enumerate}

Rheological phase transitions are fascinating novel transitions, and
the exploration of their characteristics provides new insights into the
less-explored realm of athermal non-equilibrium phase
transitions~\cite{hinrischen_2000, racz_2002}. Compared to other
well-established equilibrium transitions, rheological phase
transitions are in their infancy. We hope that our framework can aid
in a better understanding of their nature.\\


{\bf Acknowledgments.} The authors thank the Korea Institute for
Advanced Study for providing computing resources (KIAS Center for
Advanced Computation - Linux cluster system) for this work, and
especially consultations from Hoyoung Kim. We appreciate enlightening
discussions with Abbas Ali Saberi, Takahiro Hatano, Peter Olsson, and
Hisao Hayakawa. This work is supported in part by the NRF grant
No. 2017R1D1A1B06035497.\\

\section{Appendix A: simulations}
\label{append:simulations}

{\bf Numerical simulations. } We perform constant volume
molecular dynamics simulations of two-dimensional frictionless
bidisperse disks. Interactions between particles are modeled by a linear
dashpot model. Two particles $i$ and $j$ of radii $R_i^a$ and $R_j^b$
(where $a,b=0$ and $1$ stand for two different radii of bidisperse
particles) at positions ${\bf r}_i$ and ${\bf r}_j$ interact when
$\xi_{ij} = R_i^a + R_j^b - r_{ij} > 0$. Here, $\xi_{ij}$ is called
the mutual compression of particles $i$ and $j$, $r_{ij} = |{\bf r}_i
- {\bf r}_j|$. The particles interact via a linear dashpot model,
$F_{ij} = Y \xi_{ij} + \gamma \frac{d\xi_{ij}}{dt}$,
where $Y$ and $\gamma$ are denoted as elastic and dissipative constants, respectively.
Throughout the study, we adopt unitary scale $Y = 1$ and $\gamma=1$,
respectively.\\

To prevent crystallization, we use a $1:1$ binary mixture of
particles where the ratio of the radii of large and small particles is
set to $R^1/R^0=1.4$. The diameter of small particles is chosen as the
unit of the length $2R^0 = 1$, and the mass of each particle is equal to its
area, $m_a = \pi [R^a]^2$.\\

Lees-Edwards boundary conditions are applied along the ${ y}$-direction.
They create a uniform overall shear rate,~$\gd$. We use LAMMPS for our
simulations. Thanks to the developer team of LAMMPS, we were provided
with a new version of LAMMPS that prevents artificial attractive forces
arising from the dashpot model. The version can be accessed via the
mailing list of LAMMPS.\\

We used several system sizes, the smallest $L = 50 $ and the largest
$L = 300 $. We change the packing fraction by changing the number of particles
$N$ via:
\begin{equation}
  N = \frac{8}{\pi}\frac{L^2}{ 1^2 + 1.4^2} \phi,
\end{equation}
where $1$ and $1.4$ are the diameters of small and large particles,
respectively. For shear rate $\gd$ in the range $10^{-4}$ and $10^{-5}$,
the total strain is $\gamma = 30 L$ and the integration time step is
$dt = 0.1$. For the next smaller decade, the integration time step is
$dt = 0.2$.\\

\section{Appendix B: scaling ansatz}
\label{append:scaling_ansatz}

Here, we explain a formalism for deriving the scaling ansatz for
a rigidity transition. The formalism in principle can be applied to any
transition that is accompanied by a diverging length scale
$\xi$. Upon approaching the dense regime, the motion of particles becomes
coordinated. This signals the growing length scale, which diverges at
the critical density $\jamm$. This divergence is described by exponent
$\nu$ via:
\begin{equation}
  \xi \propto \delta \phi^{-\nu}.
  \label{eq:dim_nu}
\end{equation}
In the proximity of a critical point, the only fundamental length
scale $b$ is the correlation length scale, $b =
\xi$. Eq.~\ref{eq:dim_nu} can be cast into a dimensionless number as
\begin{equation}
  \Pi_{\phi} = \delta\phi b ^{1/\nu}.
  \label{eq:dimless_nu}
\end{equation}

The critical point is at $\delta\phi = 0$ and $\gd \to 0$, therefore
at $\delta\phi = 0$, the correlation length diverges upon decreasing
the shear rate:
\begin{equation}
  \xi \propto \gd ^ {-1/z},
  \label{eq:dim_z}
\end{equation}
where $z$ is the dynamic exponent. This equation can be similarly cast
into another dimensionless number via
\begin{equation}
  \Pi_{\gd} = \gd b ^z .
\end{equation}

Now, any physical quantity such the shear stress $\sxy$ also scales
with the distance from jamming $\sxy \propto \delta\phi^y$ at $\gd \to
0$. Combining this relation with Eq.~\ref{eq:dim_nu} gives:
\begin{equation}
  \sxy \propto b ^{-y/\nu},
\end{equation}
which provides the dimensionless number for this quantity
\begin{equation}
  \Pi_{\sxy} = \sxy b ^{y/\nu}.
\end{equation}

Since $\sxy$ depends on both $\delta\phi$ and $\gd$
\begin{equation}
  \Pi_{\sxy} =\mathbb{F}_0\left(\Pi_{\delta\phi}, \Pi_{\gd} \right),
\end{equation}
which results in
\begin{equation}
  \sxy b ^{y/\nu} = \mathbb{F}_0\left( \delta\phi b ^{1/\nu}, \gd b ^z\right),
  \label{eq:dimless_eq_stat}
\end{equation}
This is the dimensionless equation of state.\\

In the renormalization group method, the domain over the correlated
particles are rescaled. After renormalization, the system becomes
smaller by a factor of $b$ and therefore $\xi^{\prime} = \xi / b$. As
a result of this, the system moves away from the critical point by
renormalization. In this process, all observables and control
parameters scale with distance from critical point $b$. Equation
\ref{eq:dimless_eq_stat} describes all such scaling behaviors. Two
approaches, the intermediate asymptotic approach described by
dimensionless numbers and the renormalization group, arrive at similar
results~\cite{goldenfeld_1989}.\\

If we choose the length scale $b$ such that $\gd b ^z = 1$, then
\begin{equation}
  \sxy = \gd ^{y/z\nu}\mathbb{F}_1\left( \frac{\delta\phi}{\gd^{1/z\nu}}\right),
  \label{eq:lead_scal_rg}
\end{equation}
which is the leading scaling term. At $\delta\phi = 0$, $\sxy\propto
\gd^q$, thus $q/y = 1/z\nu$. This equation describes $\sxy$
infinitesimally close to the critical point at $\delta\phi = 0$ and
$\gd = 0$.\\

The jamming point is characterized by two principal directions given
by $\delta\phi$ and $\gd$. Each direction is accompanied by a
principal exponent: $y$ and $q$. Near the critical point only these
relevant quantities affects the dynamics. However, off the critical
point, some irrelevant parameters, $w$, may affect the dynamics. Since
this quantity is irrelevant, one cannot bring the system into the
critical point by varying such a quantity. This means that the
correlation length does not diverge if $w\to 0$. However, it may
retain a scaling form near the critical region
\begin{equation}
  \xi \propto w ^{1/\omega},
\end{equation}
which results to
\begin{equation}
  \Pi_{w} = w b ^{-\omega}.
\end{equation}

Inserting this dimensionless number into Eq.~\ref{eq:dimless_eq_stat} results in
\begin{equation}
  \sxy b ^{y/\nu} = \mathcal{F}_0\left( \delta\phi b ^{1/\nu}, \gd b ^z,
  w b ^{-\omega} \right).
\end{equation}

With $ \gd b ^z = 1$, we arrive at

\begin{equation}
  \sxy = \gd ^{y/z\nu}\mathcal{F}_1\left(
  \frac{\delta\phi}{\gd^{1/z\nu}}, w\gd^{\omega/z} \right),
\end{equation}
A Taylor expansion of this equation to the first order gives:
\begin{equation}
  \sxy = \gd ^{y/z\nu}\left[\mathcal{F}_1^{(0)}\left(
    \frac{\delta\phi}{\gd^{1/z\nu}}\right) + \gd^{\omega/z}
    \mathcal{F}_1^{(1)}\left( \frac{\delta\phi}{\gd^{1/z\nu}}\right)
    \right],
\end{equation}
This equation describes the leading correction-to-scaling term. At
$\delta\phi = 0$
\begin{equation}
  \sxy = \gd ^{y/z\nu}\left[ c_1 + c_2 \gd^{\omega/z} \right].
  \label{eq:corr_scal_lead_term}
\end{equation}

Eq.~\ref{eq:corr_scal_lead_term} can be used for scalings of the flow
curve at $\jamm$.





%

\end{document}